\documentclass[a4paper,10pt]{article}

\usepackage{publication}
\usepackage[backend=biber]{biblatex}
\usepackage{graphicx}
\usepackage{amsmath}  
\DeclareUnicodeCharacter{2009}{\,} 
\usepackage{publication}
\usepackage[backend=biber]{biblatex}
\usepackage{pdfpages}

\title{Toward Enhanced Inertial Sensing via Dynamically Soft Topological States in Piezoelectric Microacoustic Metamaterials}

\author[1]{Onurcan Kaya}
\author[1]{Niccolò Scalise Pantuso}
\author[1]{Marco Galli}
\author[2]{Jacopo M. De Ponti}
\author[1]{Tommaso Maggioli}
\author[1]{Davide Pavesi}
\author[1]{Siddhartha Ghosh}
\author[2]{Attilio Frangi}
\author[1]{Luca Colombo}
\author[1]{Benyamin Davaji}
\author[1]{Matteo Rinaldi}
\author[1]{David Horsley}
\author[1]{Cristian Cassella*}
\contact{c.cassella@northeastern.edu}

\affil[1]{Institute for NanoSystems Innovation, Northeastern University, MA, USA}
\affil[2]{Department of Civil and Environmental Engineering, Politecnico di Milano, Milano, Italy}

\date{\today}

\begin{document}

\maketitle

\begin{abstract}
    In recent decades, microelectromechanical systems (MEMS)-based gyroscopes have been employed to meet positioning and navigation demands of a plethora of commercially available devices. Most of such gyroscopes rely on electrostatic actuators with nanometer-scale air gaps—an architecture that enables large particle velocities in a proof mass and, consequently, high Coriolis-force sensitivity to angular velocity—but is inherently susceptible to damage under shock and vibration. This vulnerability is typically mitigated by purposely reducing gyroscopic sensitivity, thereby compromising readout accuracy. Microacoustic gyroscopes, by contrast, offer greater resilience to shock and vibration but currently exhibit significantly lower sensitivities. This limitation stems from the low dynamic compliance of the modes they employ—typically Lamb or Rayleigh modes—which restricts their maximum achievable particle velocity. This work presents a piezoelectric microacoustic device that overcomes this fundamental constraint by harnessing a topological interface state at the boundary between two microscale metamaterial structures. We theoretically and experimentally show that this state exhibits much higher modal compliance than Lamb or Rayleigh modes. This enables record-high particle velocities ($>$51 m/s) never reached, due to material limits, by any previously demonstrated piezoelectric gyroscope.
\end{abstract}




\section{Introduction}

\begin{figure}[!t]
  \includegraphics[width=\linewidth]{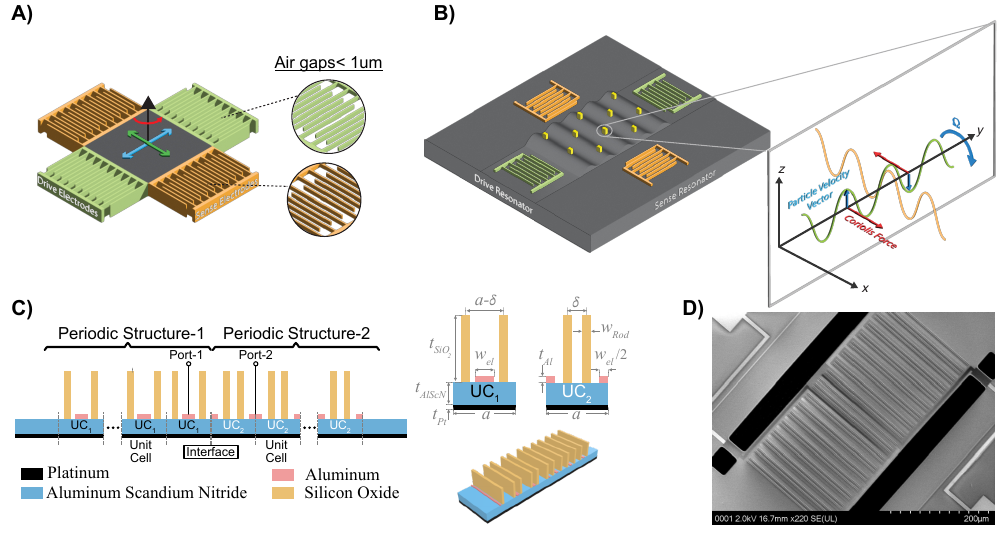}
  \caption{Conventional MEMS gyroscopic devices and the reported topological device: (a) Schematic illustration of a conventional MEMS capacitive vibratory rate gyroscope (CVRG) having a comb-drive actuator with sub-micron air gaps; (b) Schematic illustration of a conventional surface acoustic wave (SAW) gyroscope consisting of a pair of drive and sense resonators, and a forest of pillars. The scale factor of SAW gyroscopes is proportional to the magnitude of the particle velocity in the pillar, which is heavily limited by the high elastic coefficients of typical SAW gyroscopes' forming layers; (c) Schematic illustration of the proposed topological device formed by combining two periodic structures, each consisting of unit cells UC\textsubscript{1} and UC\textsubscript{2} with identical dispersion characteristics but different Zak phases; (d) Scanning electron microscopy (SEM) image of the reported topological device.}
  \label{fig1}
\end{figure}

In today’s world, microelectromechanical (MEMS) gyroscopes find use in many applications, from consumer electronics and automobiles to aerospace and military systems \cite{Market_1_acar_environmentally_2009,Market_2_barbour_inertial_2001, Market_3_challoner_boeing_2014,Market_4_geiger_mems_2008,Market_5_lane_survey_2010,Market_6_yazdi_micromachined_1998}. Currently, most MEMS gyroscopes are capacitive vibratory rate gyroscopes (CVRGs) \cite{CVRG_1_lee_surfacebulk_2000,CVRG_2_alper_single-crystal_2005,CVRG_3_zotov_self-calibrated_2014,CVRG_4_su_silicon_2014,CVRG_5_ahn_encapsulated_2014,CVRG_6_weinberg_how_2015,CVRG_7_nitzan_self-induced_2015,CVRG_8_wen_resonant_2017, CVRG_9_zhou_mitigating_2017, CVRG_10_zega_new_2018, CVRG_11_li_004_2018,CVRG_12_wu_effects_2021,CVRG_13_lysenko_analysis_2021,CVRG_14_ren_automatic_2024}. CVRGs’ operation relies on electrostatic transducers to generate a particle velocity, generally in the order of 1 m/s or less, in a proof mass. This allows generation of a Coriolis force when the proof mass is subjected to rotation – a force that CVRGs sense to quantify the angular velocity of their host platform \cite{Coriolis_1_shkel_two_2005}. The scale factor of CVRGs, which quantifies the sensitivity of their read out signal's amplitude to angular velocity, is directly proportional to the particle velocity transduced in the proof mass. Achieving a large particle velocity in the proof mass requires being able to generate large mechanical forces from CVRGs’ driving voltage, which is typically limited to a few volts in available inertial units. This is typically accomplished by scaling the size of the air-gaps between CVRGs’ electrodes down to hundreds of nanometers or less in order to maximize CVRGs’ transduction efficiency \cite{CVRG_nm_1_legtenberg_comb-drive_1996}. Unfortunately, using such small gaps (Figure \ref{fig1}a) makes CVRGs susceptible to mechanical instabilities even for moderate displacement values, as well as to catastrophic failures when subjected to high shock and vibration \cite{Reliability_1_li_shock_2014,Reliability_2_gill_review_2022,Reliability_3_liang_adhesion_2009,Reliability_4_wang_mems_2022,Reliability_5_tanner_mems_2000,Reliability_6_huang_mems_2012,Reliability_7_merlijn_van_spengen_mems_2003}. In other words, there exists a fundamental limitation in CVRGs that prevents them from being able to achieve high particle velocity in the proof mass without compromising their reliability and their resilience to shock and vibration. Because of this limitation, a significant research effort has been made in the last few years to develop alternative gyroscopic technologies not requiring air-gaps, hence not prone to instabilities and robust against shock and vibration. This research has led to the development of various piezoelectric microacoustic gyroscopes (PMGs) \cite{PMG_1_liu_4h_2024,PMG_2_tabrizian_high-frequency_2013,PMG_3_serrano_substrate-decoupled_2016,PMG_4_lee_enhancing_2017,PMG_5_davaji_towards_2017, PMG_6_pinrod_coexisting_2018, PMG_7_hodjat-shamami_eigenmode_2020, PMG_8_liu_high-q_2022, PMG_9_erturk_self-aligned_2023, PMG_10_liu_ta_2024, PMG_11_tian_toroidal_2024, PMG_12_qi_bridging_2024}. Unlike CVRGs, which typically operate at kHz frequencies, PMGs rely on the excitation of radio frequency (RF) acoustic waves – either surface acoustic waves (SAW) or bulk acoustic waves (BAW) – in thin piezoelectric plates. PMGs have a similar footprint as CVRGs. Also, they can operate without air-gaps – a feature that makes them inherently more robust and more resilient to shock and vibration than CVRGs (Figure \ref{fig1}b) \cite{PMG_3_serrano_substrate-decoupled_2016, PMG_5_davaji_towards_2017, PMG_6_pinrod_coexisting_2018, PMG_7_hodjat-shamami_eigenmode_2020}. Moreover, PMGs don’t require a hermetic encapsulation, which reduces fabrication costs and complexity. Despite these advantages, PMGs have historically exhibited lower scale factors than CVRGs. This is because Lamb waves and Rayleigh waves transduced in a piezoelectric film exhibit a low modal compliance due to the high elastic coefficients of their forming layers \cite{MatProp_AlN_Kazan_Elastic,MatProp_AlScN_wu_characterization_2018, MatProp_LNB_ogi_acoustic_2002}. Such low compliance heavily limits the maximum particle velocity ($v_{max}$) that can be generated in PMGs for any achievable quality factor value. In fact, $v_{max}$ cannot be increased indefinitely by increasing PMGs' driving power because of nonlinearities. Nonlinearities in PMGs cause a saturation of $v_{max}$ as the drive power is increased. Two main nonlinear mechanisms can affect piezoelectric devices: mechanical nonlinearities and thermal nonlinearities \cite{Nonlin_2_givois_dynamics_2021}. Thermal nonlinearities represent the main source of nonlinearities for most piezoelectric devices operating in the RF range \cite{Nonlin_1_tazzoli_experimental_2012,Nonlin_3_miller_nonlinear_2014,Nonlin_4_segovia-fernandez_thermal_2013,Nonlin_5_wang_piezoelectric_2015,Nonlin_6_lu_study_2015}. Beyond the impact of nonlinearities, driving PMGs with excessive power can also cause catastrophic failures. Such failures can originate from crack formation and propagation or from damages in the electrodes due to acoustomigration or electromigration \cite{PowerHand_1_wunnicke_thermal_2009,PowerHand_2_larson_power_2000,PowerHand_3_van_der_wel_thermal_2009}. All these limitations explain why prior PMGs could not achieve velocities higher than a few meters per second  \cite{PMG_10_liu_ta_2024}. Alternative gyroscopic designs are also available \cite{PMG_1_liu_4h_2024,PMG_3_serrano_substrate-decoupled_2016,PMG_7_hodjat-shamami_eigenmode_2020,PMG_8_liu_high-q_2022,PMG_10_liu_ta_2024, PMG_12_qi_bridging_2024}, using a piezoelectric actuation for placing the proof mass into motion and an electrostatic transducer for readout. Using the electrostatic readout enables higher scale factors than conventional PMGs. However, like CVRGs, these $hybrid$ gyroscopes recur to sub-micron capacitive air gaps to maximize the achievable scale factor, which lowers the resilience to shock and vibration. 

In order to overcome the limitation of the existing MEMS gyroscopes, research efforts have been made to develop photonic gyroscopic devices. These devices, mostly based on the Sagnac effect, offer exceptional scale factors, along with inherent resilience to shock and vibration since they lack moving parts \cite{Optical_1_wu_silicon_2018, Optical_2_wu_mode-assisted_2019}. Yet, their large footprint, high power consumption, and high cost restrict their use to very specialized applications \cite{PMG_1_liu_4h_2024}.

On a separate note, researchers across several disciplines – including photonics, acoustics, and mechanics – have started emulating interactions discovered in condensed matter physics using classical waves in periodic media. These efforts have led to the practical realization of photonic, acoustic, and mechanical metamaterial devices exhibiting exotic properties, such as negative refraction, negative effective constitutive parameters, and non-reciprocity \cite{MetaGen_1_guild_plasmonic-type_2012,MetaGen_2_zhu_holey-structured_2011,MetaGen_3_tuniz_metamaterial_2013,MetaGen_4_zigoneanu_three-dimensional_2014, MetaGen_5_moleron_acoustic_2015,MetaGen_6_zangeneh-nejad_analogue_2021,MetaGen_7_zanotto_metamaterial-enabled_2022}. These devices have enabled functionalities that could not be attained using conventional wave-propagating structures, such as cloaking and imaging beyond the limit of diffraction \cite{MetaGen_3_tuniz_metamaterial_2013}. Some of these devices have been able to reproduce properties of topological quantum systems, such as the quantum Hall effect, where time-reversal symmetry is broken through applied external fields, or the quantum Spin Hall effect, in which symmetry breaking is achieved through spin-orbit interactions \cite{MetaInv_1_xiao_geometric_2015,MetaInv_2_fu_topological_2007}. This has led to the demonstration of a new class of waves offering unique advantages over conventional ones, including robustness against fabrication imperfections and structural disorder, immunity to back-scattering during propagation, and strong mode localization \cite{MetaInvAppln_1_de_ponti_localized_2024,MetaInvAppln_2_xi_soft-clamped_2025,MetaInvAppln_3_prabith_nonlinear_2024,MetaInvAppln_4_lu_observation_2017,MetaInvAppln_5_mousavi_topologically_2015}. In this context, our group has recently demonstrated a piezoelectric RF MEMS resonator relying on microacoustic periodic structures to transduce topological interface states – a feature that enables both high responsivity and low thermomechanical noise when using such device for monitoring localized parameters \cite{MetaInvAppln_1_de_ponti_localized_2024}. Despite this important milestone, the properties of topological interface states in MEMS devices remain largely unexplored, as does their potential to serve as vibration modes in gyroscopic systems. 

In this Article, we present a piezoelectric microacoustic device that excites a topological interface state (IS) and shows, both theoretically and experimentally, that such a mode of vibration enables a significantly higher particle velocity than the typical modes harnessed by the available PMGs. Using a one-dimensional mass–spring model and finite element modeling, we demonstrate that ISs exhibit significantly higher modal compliance compared to any other trivial (e.g., not topological) mode that can be excited in the same structure. Such $dynamic$ $softness$ is the key to achieving much higher particle velocities before the impact of nonlinearities becomes overwhelming or a catastrophic fracture occurs. By leveraging this dynamic softness, the reported device has been able to reach a measured particle velocity of 51.3 m/s. Such high particle velocity is the highest particle velocity ever reported in MEMS devices. Our findings create new means to surpass a fundamental trade-off in MEMS gyroscopes, simultaneously enabling the typical robustness to shock and vibration of PMGs and a sensitivity to angular rotation greatly enhanced by the exotic characteristics of topological ISs.

\section{Design and Modeling}

The reported device is formed by connecting two periodic structures (labeled “Periodic Structure-1” (PS-1) and “Periodic Structure-2 (PS-2), see Figure \ref{fig1}c) reconstructing the Su-Schrieffer-Heeger (SSH) model \cite{SSH_PhysRevLett.42.1698} in an elastic framework at the microscale. The two periodic structures are characterized by different unit cells, labeled UC\textsubscript{1} and UC\textsubscript{2} respectively, and each structure consists of a chain of nine unit cells. Both UC\textsubscript{1} and UC\textsubscript{2} are formed by a thin layer of Aluminum Scandium Nitride (AlScN) – a piezoelectric material commonly used to build microacoustic RF filters \cite{AlScN_1_zukauskaite_editorial_2023} – deposited on top of an unpatterned platinum (Pt) layer. The AlScN layer hosts a set of silicon dioxide (SiO\textsubscript{2}) rods (two per unit cell), along with a series of metallic strips made of aluminum (Al). All these strips are electrically grounded, except for two metal strips near the shared interface between the periodic structures. Such two strips, forming two electrical ports (Port-1 and Port-2), are respectively used to excite the device and to sense its piezo-generated charge. UC\textsubscript{1} and UC\textsubscript{2} have the same unit cell size ($a$) but their SiO\textsubscript{2} rods are spaced by $a-\delta$ and $\delta$,  respectively (see Supporting Information Section S1 for all geometric parameters of the reported device). UC\textsubscript{1} and UC\textsubscript{2} show identical elastic dispersion curves, including the existence of an overlapping band gap. However, they exhibit different values of Zak phase, which is a well-established topological invariant serving as an indicator of band inversion \cite{MetaInv_1_xiao_geometric_2015,MetaInv_3_zak_berrys_1989}. Connecting PS-1 to PS-2 allows to reconstruct the SSH model by creating an interface with broken periodicity (i.e., a topological interface) at the shared boundary of the two periodic structures, allowing the emergence of topologically protected interface states (ISs). Such ISs are localized at the topological interface and their dispersion curves lie within the bandgap of the periodic structure.

\begin{figure}[!t]
  \includegraphics[width=\linewidth]{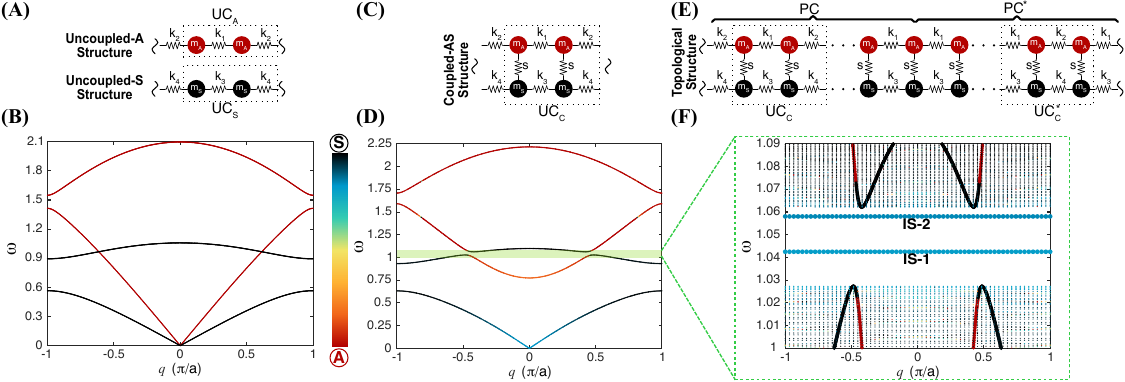}
  \caption{Verification of the existence of interface states via a toy model: (a) Unit cells of the uncoupled antisymmetric (uncoupled-A) and uncoupled symmetric (uncoupled-S) structures representing the propagation of uncoupled antisymmetric and symmetric waves; (b) Unit cell dispersion curves of the uncoupled structures; (c) The coupling between antisymmetric and symmetric waves is introduced via an interconnecting spring, $s$, forming a coupled unit cell (UC\textsubscript{C}) of a periodic coupled structure, denoted as coupled antisymmetric-symmetric (coupled-AS) structure; (d) Unit cell dispersion analysis of UC\textsubscript{C} showing the opening of a complete band gap; (e) Topological structure formed by combining two chains consisting of two different unit cells; (f) Dispersion curves for the topological structure, revealing two interface states inside the same band gap shown in (d).}
  \label{fig2}
\end{figure}

The scanning electron microscopy (SEM) image of the fabricated topological device is shown in Figure \ref{fig1}d. The device reported here shows two ISs (IS-1 and IS-2) that emerge from the interaction of an antisymmetric (A0) and symmetric (S0) Lamb modes. Both ISs are strongly localized near the common boundary of the two periodic structures but have different predominant polarizations (e.g., one is mainly longitudinal and the other one is mainly flexural). The emergence of the ISs and their main properties can be studied by recurring to three toy models consisting of specific sets of masses and springs capturing the propagation of symmetric and antisymmetric modes of vibration across the periodic structures for three distinct cases. The first case assumes no coupling between symmetric and antisymmetric modes. Figure \ref{fig2}a shows the equivalent unit cells, namely UC\textsubscript{A} and UC\textsubscript{S}, corresponding to two uncoupled periodic structures, denoted as uncoupled antisymmetric (uncoupled-A) structure and uncoupled symmetric (uncoupled-S) structure. Each uncoupled structure consists of identical masses (m\textsubscript{A} and m\textsubscript{S}, respectively) and alternating spring pairs (k\textsubscript{1}-k\textsubscript{2} and k\textsubscript{3}-k\textsubscript{4}, respectively). The unit cell's dispersion curves for these uncoupled structures are given in Figure \ref{fig2}b. The curves have been extracted by imposing Bloch-Floquet periodic boundary conditions \cite{Floquet_1_gomez_garcia_floquet-bloch_2015}. The second case introduces coupling between the symmetric and antisymmetric modes via an interconnecting spring, s, between UC\textsubscript{A} and UC\textsubscript{S}, forming a coupled unit cell (UC\textsubscript{C}) of a periodic coupled structure, denoted as coupled antisymmetric-symmetric (coupled-AS) structure, as illustrated in Figure \ref{fig2}c. Introducing this coupling term leads to a change in the dispersion curves of antisymmetric and symmetric modes, including the opening of a complete band gap (see Figure \ref{fig2}d). The third and last case examines a structure with broken periodicity. This structure is formed by combining two periodic chains (PCs), denoted as PC and PC\textsuperscript{*}. PC consists of a series of UC\textsubscript{c}, while PC\textsuperscript{*} incorporates an inverted coupled unit cell (UC\textsubscript{c}\textsuperscript{*}) configuration. UC\textsubscript{c} is characterized by a set of intra-cell and inter-cell spring constants (k\textsubscript{1}-k\textsubscript{3} and k\textsubscript{2}-k\textsubscript{4}, respectively), while UC\textsubscript{c}\textsuperscript{*} uses intra- and inter-cell spring constants equal to k\textsubscript{2}-k\textsubscript{4} and k\textsubscript{1}-k\textsubscript{3} respectively. Moving forward, we will refer to the structure shown in Figure \ref{fig2}e as the "topological structure". We can confirm the presence of interface states in the topological structure by recurring to a supercell analysis \cite{SuperCell_1_zhang_topological_2022,SuperCell_2_chu_temperature_2024,Supercell_3_rajabpoor_alisepahi_breakdown_2023}. This analysis allows to extract the dispersion of the supercell, shown in Figure \ref{fig2}f, for frequencies around the band gap of UC\textsubscript{C}. As expected, Figure \ref{fig2}f reveals the existence of two ISs inside the band gap. For clarity, we replotted the dispersion curves of UC\textsubscript{C} in Figure \ref{fig2}f, together with the supercell's dispersion curves for the topological structure. This was done to highlight how the ISs found through our supercell analysis fall within the same band gap identified for the structure in Figure \ref{fig2}c. More information about the derivation of the dispersion curves for Figure \ref{fig2}a,c,e is provided in Supporting Information Section S2. It is also worth emphasizing that the ISs found through our supercell analysis are topologically protected from defects. This is ensured by the fact that the Zak phase of UC\textsubscript{c} and UC\textsubscript{c}\textsuperscript{*} are different (see Supporting Information Section S3 for more information on the calculation of the Zak phase for this case).

\begin{figure}[!t]
  \includegraphics[width=\linewidth]{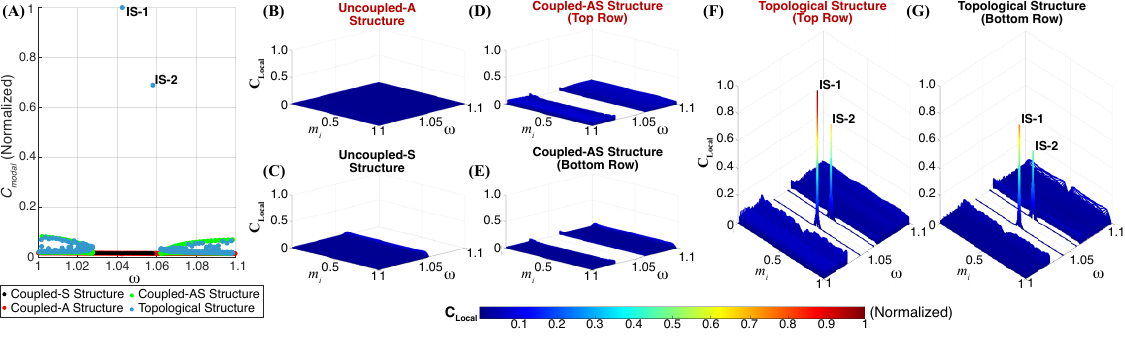}
  \caption{Comparison of topological and trivial modes via modal and dynamic local compliances of the mass-spring toy models: (a) Normalized Modal compliance ($C_{modal}$) of uncoupled, coupled, and topological structures, demonstrating that ISs exhibit much higher $C_{modal}$ than any other modes of uncoupled and coupled structures; (b-g) Dynamic local compliance ($C_{Local}$) of uncoupled-A (b), uncoupled-S (c), coupled-AS (d-e), and topological (f-g) structures, showing that masses of the topological structure located near the interface between PC and PC\textsuperscript{*} have significantly higher compliance values compared to trivial modes.}
  \label{fig3}
\end{figure}

After confirming the existence of the ISs, we used the toy models described in Figures \ref{fig2}a, \ref{fig2}c, and \ref{fig2}e to compare the maximum achievable particle velocity of the ISs with that of symmetric and antisymmetric modes for both uncoupled and coupled cases. For this analysis, we used stress-free boundary conditions rather than the Floquet boundary conditions needed for running a dispersion analysis. Moreover, the number of unit cells for each structure was adjusted such that all structures had the same total mass (See Supporting Information Table S4). This ensures that differences in particle velocities arise solely from changes in the modal characteristics. After performing an eigen-study for each of these structures, we calculated the modal compliance ($C_{modal}$ ) for each identified eigenmode.

\begin{equation}
\mathcal{C}_{\mathrm{modal}} = \frac{1}{k_{\mathrm{eff}}} = \frac{1}{\omega_{\mathrm{res}}^2 m_{\mathrm{eff}}}
\label{eqn_Cmodal}
\end{equation}

$C_{modal}$ is the inverse of the effective modal stiffness ($k_{eff}$). It can be extracted, for each eigenmode, by using the natural resonance frequency ($\omega_{res}$) and the effective mass ($m_{eff}$) of the eigenmode (See  Equation \ref{eqn_Cmodal} and Supporting Information Sections S4.1 and S4.2 for more information about the calculation of $C_{modal}$). The calculated $C_{modal}$ values were normalized with respect to the maximum $C_{modal}$ obtained across all structures and are given in Figure \ref{fig3}a. Figure \ref{fig3}a reveals that the ISs of the topological structure can exhibit much higher $C_{modal}$ than any other modes of the uncoupled and coupled structures. This is due to ISs' superior mode localization. Such mode localization allows the ISs to show a significantly lower $m_{eff}$ compared to any other mode of the structures under study. This was confirmed by running an additional study, aiming to extract the dynamic local compliance ($\mathbf{C}_{\text{Local}}$) for each mass forming the uncoupled (see Figures \ref{fig3}b and \ref{fig3}c), the coupled (see Figures \ref{fig3}d and \ref{fig3}e) and the topological structures (see Figures \ref{fig3}f and \ref{fig3}g). $\mathbf{C}_{\text{Local}}$ is a $N$x$M$ matrix where, for each structure under study, $N$ is equal to the number of masses and $M$ corresponds to the number of angular frequencies ($\omega$) analyzed during the study. In particular, each component of $\mathbf{C}_{\text{Local}}$ corresponds to the displacement of a specific mass of the analyzed structure (when assuming a fixed arbitrary quality factor, Q) when subjected to a unitary force at a specific angular frequency corresponding to one of the structure's angular natural frequencies. More information about $\mathbf{C}_{\text{Local}}$ and its derivation is available in Supporting Information Sections S4.1 and S4.3 \cite{CLocal_1_mcconnell_modal_2001}. The calculated components of $\mathbf{C}_{\text{Local}}$ values were normalized with respect to the maximum dynamic local compliance value across all cases and are presented in Figures \ref{fig3}b---\ref{fig3}g. Such values are plotted against mass index ($m_i$), normalized by the total number of masses in the corresponding structure, and $\omega$. Evidently, $\mathbf{C}_{\text{Local}}$ for both the uncoupled and coupled structures exhibits a uniform spatial profile along the chain of masses. In contrast,  $\mathbf{C}_{\text{Local}}$ for the topological structure (Figures \ref{fig3}f and \ref{fig3}g) shows sharp and significant peaks in the proximity of the mass located at the interface between PC and PC\textsuperscript{*} and at the natural resonance frequencies of IS-1 and IS-2. This confirms that the increase of the $C_{modal}$ observed in Figure \ref{fig3}a originates from the ability of the masses near the interface between PC and PC\textsuperscript{*} to displace significantly more, when a force is applied at the resonance frequency of IS-1 or IS-2, than any mass included in all the analyzed structures when applying the same force at a different frequency coinciding with the resonance frequency of a trivial mode. This ability is the key to achieving much higher particle velocities at the interface between PC and PC\textsuperscript{*} than possible when exploiting conventional modes. These results theoretically demonstrate the capability of ISs to outperform the maximum velocity achievable by conventional symmetric and antisymmetric modes for any given input driving voltage.

\begin{figure}[!t]
  \includegraphics[width=\linewidth]{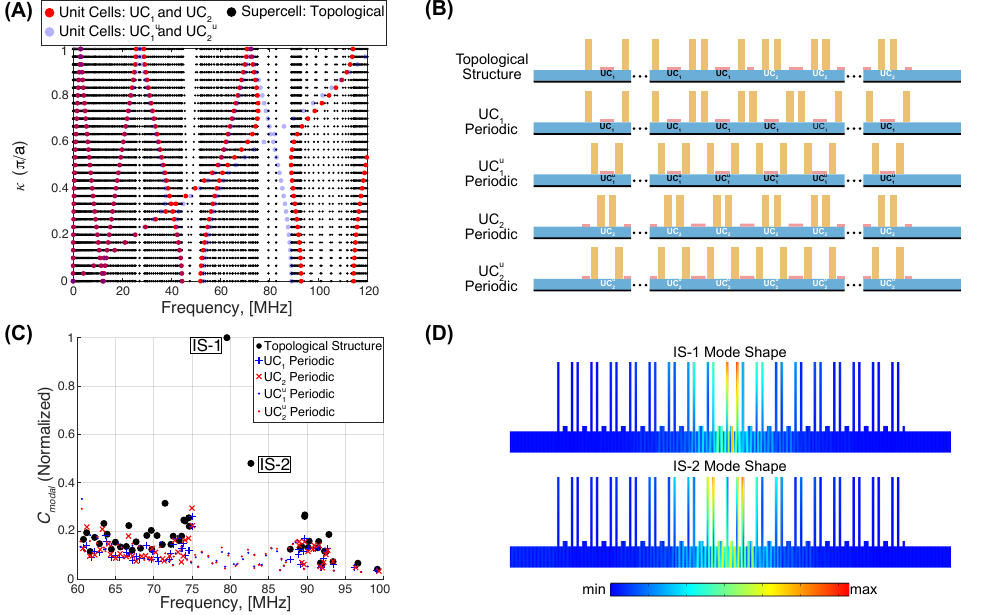}
  \caption{Evidence of the existence of the reported device's ISs through FEM simulations: (a) Dispersion curves showing that unit cells of the topological structure possess a band gap within which topological states exist. This band gap disappears when rods are distributed uniformly; (b) Structures used in modal compliance ($C_{modal}$) analysis to compare the particle velocity performance of ISs with that of trivial modes; (c) Resulting $C_{modal}$ values around the band gap showing that ISs have higher $C_{modal}$ than any other trivial mode; (d) Total displacement mode shapes of IS-1 and IS-2, showing strong mode localization.}
  \label{fig4}
\end{figure}

To further confirm the validity of our analytical findings when considering the full spectrum of modes that exist in the stuck of our reported device, we performed an additional study using Finite Element Methods (FEM). First, we analyzed the unit cell’s elastic dispersion characteristics of four structures: (i-ii) UC\textsubscript{1} and UC\textsubscript{2}, and (iii-iv) modified versions of UC\textsubscript{1} and UC\textsubscript{2} with uniform rod spacing ($\delta$ = a/2), denoted as UC\textsubscript{1}\textsuperscript{u} and UC\textsubscript{2}\textsuperscript{u}, respectively. The resulting dispersion curves were plotted in terms of corresponding mode’s lateral wavevector ($\kappa$) vs frequency and shown in Figure \ref{fig4}a, revealing a band gap for UC\textsubscript{1} and UC\textsubscript{2}, whereas no such band gap exists for UC\textsubscript{1}\textsuperscript{u} and UC\textsubscript{2}\textsuperscript{u}. Subsequently, we analyzed the supercell dispersion of the topological structure formed by connecting a chain of UC\textsubscript{1} with a chain of UC\textsubscript{2}. Details about the methodology used to extract the dispersion characteristics of these geometries are provided in Supporting Information Section S5.1. Although UC\textsubscript{1} and UC\textsubscript{2} have identical dispersion curves, they possess opposite Zak phases (See Supporting Information Section S5.2), which enables the emergence of two topological ISs at the topological interface shared by the cascaded chains. Next, we performed another finite element study to compare ISs' ability to achieve high particle velocities to that of any other mode. For this analysis, as depicted in Figure \ref{fig4}b, we considered five structures with an identical total number of unit cells: (i) the reported topological device, (ii) the periodic-UC\textsubscript{1} device,  a periodic structure formed by a chain of UC\textsubscript{1}, (iii) the periodic-UC\textsubscript{1}\textsuperscript{u} device, a periodic structure formed by a chain of UC\textsubscript{1}\textsuperscript{u}, (iv) the periodic-UC\textsubscript{2} device, a periodic structure formed by a chain of UC\textsubscript{2}, and (v) the periodic-UC\textsubscript{2}\textsuperscript{u}, a periodic structure formed by a chain of UC\textsubscript{2}\textsuperscript{u}. Following what we did to analyze the mass-spring model, we extracted $C_{modal}$ for all eigenmodes of the considered geometries. $C_{modal}$ was then normalized with respect to the maximum observed $C_{modal}$ value across all cases and plotted in Figure \ref{fig4}c. An additional study that compares the particle velocity performance of Rayleigh modes and ISs is provided in Supporting Information Section S6. Evidently, our FEM results confirm that ISs exhibit significantly higher $C_{modal}$ values than any other trivial mode. As discussed earlier, ISs' high $C_{modal}$ value results from their strong mode localization, which has been confirmed here through FEM in Figure \ref{fig4}d. It is worth mentioning that reconstructing the SSH model to generate topological ISs is not the only way to achieve localization. For instance, localized modes that are not topological can be created through structural defects in a periodic structure \cite{defect_1_jo_revealing_2022,defect_2_jo_double_2021}. This makes it important to assess the advantage of the topological ISs transduced by our reported device over defect modes. As detailed in Supporting Information Section S7, defect modes exhibit lower particle velocities than ISs, which further reduce when considering local temperature changes  around the defects caused by self-heating. In this regard, the strong sensitivity of defect modes to local temperature variations makes the adoption of defect modes not ideal when trying to reach the highest possible particle velocity by increasing the adopted driving power up to the maximum sustainable value. 

\section{Measurement Results}

\begin{figure}[!b]
  \includegraphics[width=\linewidth]{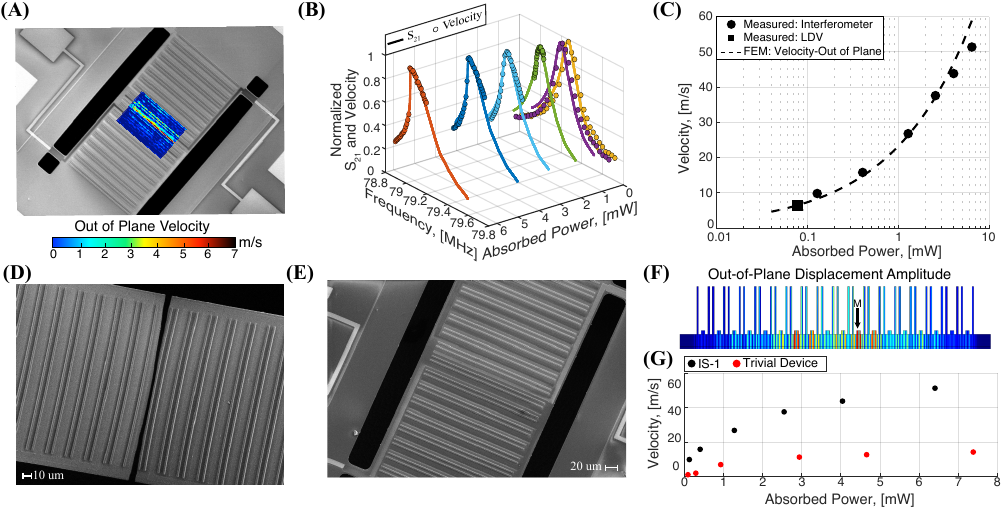}
  \caption{Experimental results: (a) Laser Doppler Vibrometer measurement of the topological structure showing the magnitude of the out-of-plane velocity around the interface between PS-1 and PS-2 for a driving power equal to 0.06 W; (b) Recorded S\textsubscript{21} and the corresponding out-of-plane velocities during interferometric measurements, each normalized to its maximum value at a given absorbed power. Both responses show a Duffing-type softening nonlinearity for power absorbed levels exceeding 4 mW; (c) Measured and FEM simulated maximum out-of-plane particle velocities at the interface between PS-1 and PS-2 under different absorbed power levels. The measured performance of the reported device is closely matching with the expected values based on FEM simulation, achieving a maximum out-of-plane velocity of 51.3 m/s under an absorbed power of 6.4 mW; (d) Scanning Electron Microscope (SEM) image of the device after being tested under 10 mW absorbed power, showing fracture at the interface between two periodic structures; (e) SEM image of the trivial device used in particle velocity performance comparison between IS-1 and the trivial mode; (f) FEM simulated out-of-plane particle velocity distribution of the trivial device, showing the location of the maximum out-of-plane particle velocity indicated by the marker M; (g) Measured maximum out-of-plane particle velocities of the topological and trivial devices, showing that the topological device exhibits higher out-of-plane particle velocities for any given absorbed power level.}
  \label{fig5}
\end{figure}

Next, we proceeded with our experimental characterization (see Supporting Information Section S8 for details of the microfabrication process). First, we confirmed the existence of our device’s ISs through electrical two-port scattering parameter (S-parameter) measurements and through an out-of-plane displacement characterization run through an in-house Laser Doppler Vibrometer (LDV). The complete S\textsubscript{21} response of the device for a frequency span including the resonance frequencies of IS-1 and IS-2 is provided in Supporting Information Section S9, together with the LDV measurements showing the out-of-plane displacement amplitudes of the device for both ISs around the interface of PS-1 and PS-2. In line with the findings from our analytical study and finite element analysis, we found IS-1 to exhibit a higher out-of-plane displacement compared to IS-2. Consequently, we focused all our experimental characterization on IS-1. As an example, Figure \ref{fig5}a shows a color map of the out-of-plane particle velocity of IS-1, measured with our LDV system under a driving power of 0.06 W at the resonance frequency of IS-1. Evidently and as expected, IS-1 is localized at the interface between PS-1 and PS-2. The applied drive power level corresponds to a power absorbed by the device equal to 0.078 mW. This value of power absorbed, like any other power absorbed values in our experimental characterization, was calculated from the driving power and from the device's S-parameters at the resonance frequency of IS-1. A discussion on the methodology used for the calculation of the power absorbed can be found in Supporting Information Section S10. The maximum out-of-plane particle velocity measured with our LDV system is 6.375 m/s. Following our LDV characterization, we used an interferometric setup to characterize the maximum out-of-plane particle velocity achieved by IS-1 for driving powers beyond the output power limit of our LDV system. During this test, the device was driven with power levels ranging from 0.01 W to 0.8 W, while the drive frequency was swept around the frequency of IS-1 using a Vector Network Analyzer (VNA) connected to an RF amplifier (see Supporting Information Section S11 for a detailed description of the test setup). These power levels correspond to powers absorbed by the device ranging from 0.127 mW to 10 mW. Figure \ref{fig5}b reports the frequency responses of the recorded S\textsubscript{21} and out-of-plane particle velocity attained during our interferometric measurements. These responses have been normalized, for each swept driving power, to their maximum values attained during the frequency sweep. Evidently, the extracted frequency responses of the S\textsubscript{21} and out-of-plane particle velocity follow closely matching trends for all levels of driving and absorbed powers. In particular, both responses show increased distortion due to softening, which causes a progressive reduction of the frequency of IS-1 (f\textsubscript{IS-1}) as the device  absorbs more power. The increased level of absorbed power, in fact, generates a local rise in temperature near the boundary between PS-1 and PS-2. This causes an increase of the modal compliance for frequencies approaching f\textsubscript{IS-1} due to the temperature coefficient of the Young's modulus of the device's forming layers \cite{TCE_2_lin_thermal_2010,TCE_3_10.1063/5.0201566}. The out-of-plane particle velocities extracted through both our interferometric and LDV systems when driving the device at a frequency equal to f\textsubscript{IS-1} and when using different levels of driving powers  are listed in Figure \ref{fig5}c in terms of the device's corresponding power absorbed values. Evidently, the reported device can reach a maximum measured out-of-plane particle velocity of 51.3 m/s for an input power of 0.5 W, corresponding to an absorbed power of 6.4 mW. It is worth mentioning that our measured particle velocities closely match the expected values based on finite element modeling (see the dashed line in Figure \ref{fig5}c), especially for absorbed power levels lower than 4 mW. Minor deviations between measured and simulated out-of-plane particle velocities are observed at absorbed power levels exceeding 4 mW. These discrepancies can be attributed to nonlinearities, which affect the device's response experimentally (see Figure \ref{fig5}b) but that our finite element tool does not consider. Further increasing the input power to 0.8 W, corresponding to an absorbed power of 10 mW, caused a catastrophic fracture of the device, as shown in the Scanning Electron Microscopy (SEM) image in Figure \ref{fig5}d. Evidently, the SEM of our device shows a nearly straight crack. This crack is located near the common interface of PS-1 and PS-2, thus in the device's region reaching the maximum out-of-plane velocity and first principal stress, as shown by the FEM-simulated frequency domain studies in Supporting Information Section 12. Finally, we compared the particle velocity achieved when using IS-1 with that attained when relying on a trivial mode. To do so, we extracted the out-of-plane particle velocity reached by another device (e.g., a "trivial device") fabricated on the same chip as our reported topological device. The SEM image of the fabricated trivial device is given in Figure 5e. The trivial device consists of a chain of only one type of unit cells (i.e., UC\textsubscript{1}\textsuperscript{u} periodic in Figure \ref{fig4}b). Also, it shows a resonance frequency at 74.3 MHz with a 3-dB quality factor (Q\textsubscript{3-dB} equal to 334) matching closely that of IS-1 (477). Further details on the geometric comparison between the reported topological device and the trivial device are provided in Supporting Information Section S13, together with a comparison of their S\textsubscript{21} responses. To run this additional test, we recurred to the same interferometric setup adopted to characterize IS-1. The measured out-of-plane particle velocities of the trivial device---taken at the location of maximum out-of-plane particle velocity predicted by our FEM simulations (Figure \ref{fig5}f)---for varying absorbed power levels are compared with those of IS-1 in Figure \ref{fig5}g. As can be seen, both particle velocity trends tend to saturate for increasing absorbed power levels. However, for the same level of absorbed power (and thus for comparable degrees of nonlinearity), the topological device achieves a maximum particle velocity nearly four times higher than that of the trivial device. This large difference in out-of-plane velocities is attributed to the higher compliance of IS-1. In fact, the trivial device exhibits a 3.9 times lower $C_{modal}$ value than IS-1 in our FEM analysis in Figure \ref{fig4}c. 

\section{Conclusion}

This work presents a novel microacoustic wave device that leverages topologically protected interface states (IS) formed by combining two periodic structures with identical dispersion characteristics but different Zak phases. Using analytical models and finite element modeling, we showed that ISs are characterized by a significantly larger modal compliance than any other Lamb or Rayleigh mode in such structures. This enables the achievement of higher particle velocities. We verified these claims experimentally by measuring the out of plane velocities of a manufactured topological device, which allowed us to reach a record-high measured velocity of 51 m/s. 

\begin{figure}[!h]
  \includegraphics[width=\linewidth]{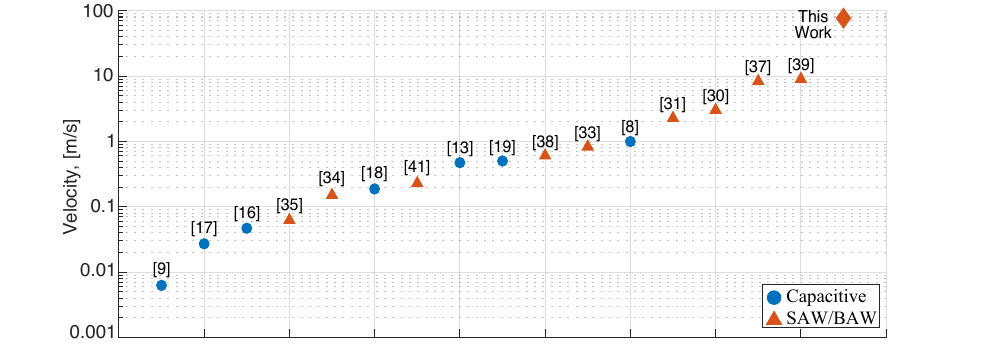}
  \caption{Comparison of reported device's particle velocity with the particle velocities previously reported in capacitive, surface acoustic wave (SAW), and bulk acoustic wave (BAW) gyroscopes.}
  \label{fig6}
\end{figure}

Figure \ref{fig6} compares the maximum measured particle velocity of the reported device with that of recent capacitive, surface acoustic wave (SAW) and BAW gyroscopes reported in the literature, demonstrating that the reported device achieves more than five times higher particle velocity than any other device ever reported. These findings suggest that recurring to topological interface states can surpass one of the biggest limitations of piezoelectric devices used for gyroscopic applications, shaping a new path to achieve MEMS gyroscopes with both a superior scale factor and high resilience to shock and vibration (see Supporting Information Section S14). In this regard, future work will focus on designing PMGs leveraging topological interface states at distinct interfaces to maximize the equivalent modal mass - a fundamental feature to maximize the Coriolis force generated by rotation - and on engineering an effective read out strategy for sensing the displacement produced by the Coriolis force with a high transduction efficiency. 

\paragraph{Supporting Information} 
The Supporting Information is provided at the end of the manuscript.

\paragraph{Acknowledgements}
This research was developed with funding from the Defense Advanced Research Projects Agency through the "Nimble Ultrafast Microsystems" (NIMBUS) program. Distribution Statement "A" (Approved for Public Release, Distribution Unlimited). The views, opinions, and/or findings expressed are those of the author(s) and should not be interpreted as representing the official views or policies of the Department of Defense or the U.S. Government. The authors would like to thank Northeastern University Kostas Cleanroom and Harvard CNS staff.



\printbibliography

@article{AlScN_1_zukauskaite_editorial_2023,
	title = {Editorial for Special Issue “Piezoelectric Aluminium Scandium Nitride ({AlScN}) Thin Films: Material Development and Applications in Microdevices”},
	volume = {14},
	rights = {http://creativecommons.org/licenses/by/3.0/},
	issn = {2072-666X},
	url = {https://www.mdpi.com/2072-666X/14/5/1067},
	doi = {10.3390/mi14051067},
	shorttitle = {Editorial for Special Issue “Piezoelectric Aluminium Scandium Nitride ({AlScN}) Thin Films},
	pages = {1067},
	number = {5},
	journaltitle = {Micromachines},
	author = {Žukauskaitė, Agnė},
	urldate = {2025-10-05},
	date = {2023-05},
	langid = {english},
	note = {Publisher: Multidisciplinary Digital Publishing Institute},
}

@article{SSH_PhysRevLett.42.1698,
  title = {Solitons in Polyacetylene},
  author = {Su, W. P. and Schrieffer, J. R. and Heeger, A. J.},
  journal = {Phys. Rev. Lett.},
  volume = {42},
  issue = {25},
  pages = {1698--1701},
  numpages = {0},
  year = {1979},
  publisher = {American Physical Society},
  doi = {10.1103/PhysRevLett.42.1698},
  url = {https://link.aps.org/doi/10.1103/PhysRevLett.42.1698}
}

@article{Floquet_1_gomez_garcia_floquet-bloch_2015,
	title = {Floquet-Bloch Theory and Its Application to the Dispersion Curves of Nonperiodic Layered Systems},
	volume = {2015},
	url = {https://onlinelibrary.wiley.com/doi/abs/10.1155/2015/475364},
	doi = {https://doi.org/10.1155/2015/475364},
	pages = {475364},
	number = {1},
	journaltitle = {Mathematical Problems in Engineering},
	author = {Gómez García, Pablo and Fernández-Álvarez, José-Paulino},
	date = {2015},
	note = {\_eprint: https://onlinelibrary.wiley.com/doi/pdf/10.1155/2015/475364},
}

@inproceedings{TCE_2_lin_thermal_2010,
	title = {Thermal compensation for aluminum nitride {Lamb} wave resonators operating at high temperature},
	url = {https://ieeexplore.ieee.org/document/5556381},
	doi = {10.1109/FREQ.2010.5556381},
	urldate = {2024-04-22},
	booktitle = {2010 {IEEE} {International} {Frequency} {Control} {Symposium}},
	author = {Lin, Chih-Ming and Yen, Ting-Ta and Felmetsger, Valery V. and Hopcroft, Matthew A. and Kuypers, Jan H. and Pisano, Albert P.},
	month = jun,
	year = {2010},
	note = {ISSN: 2327-1949},
	pages = {14--18},
}

@article{TCE_3_10.1063/5.0201566,
    author = {Sui, Wen and Pearton, Stephen J. and Feng, Philip X.-L.},
    title = {A review on temperature coefficient of frequency (TCf) in resonant microelectromechanical systems (MEMS)},
    journal = {Applied Physics Reviews},
    volume = {12},
    number = {2},
    pages = {021330},
    year = {2025},
    month = {06},
    issn = {1931-9401},
    doi = {10.1063/5.0201566},
    url = {https://doi.org/10.1063/5.0201566},
    eprint = {https://pubs.aip.org/aip/apr/article-pdf/doi/10.1063/5.0201566/20568297/021330_1_5.0201566.pdf},
}

@article{CLocal_1_mcconnell_modal_2001,
	title = {Modal Testing},
	volume = {359},
	issn = {1364-503X},
	url = {https://www.jstor.org/stable/3066391},
	pages = {11--28},
	number = {1778},
	journaltitle = {Philosophical Transactions: Mathematical, Physical and Engineering Sciences},
	author = {{McConnell}, Kenneth G.},
	urldate = {2025-10-12},
	date = {2001},
	note = {Publisher: The Royal Society},
}

@article{defect_1_jo_revealing_2022,
	title = {Revealing defect-mode-enabled energy localization mechanisms of a one-dimensional phononic crystal},
	volume = {215},
	issn = {0020-7403},
	url = {https://www.sciencedirect.com/science/article/pii/S0020740321006603},
	doi = {10.1016/j.ijmecsci.2021.106950},
	pages = {106950},
	journaltitle = {International Journal of Mechanical Sciences},
	shortjournal = {International Journal of Mechanical Sciences},
	author = {Jo, Soo-Ho and Yoon, Heonjun and Shin, Yong Chang and Youn, Byeng D.},
	urldate = {2025-10-06},
	date = {2022-02-01},
}

@article{defect_2_jo_double_2021,
	title = {Double defects-induced elastic wave coupling and energy localization in a phononic crystal},
	volume = {8},
	issn = {2196-5404},
	url = {https://doi.org/10.1186/s40580-021-00277-4},
	doi = {10.1186/s40580-021-00277-4},
	pages = {27},
	number = {1},
	journaltitle = {Nano Convergence},
	shortjournal = {Nano Convergence},
	author = {Jo, Soo-Ho and Shin, Yong Chang and Choi, Wonjae and Yoon, Heonjun and Youn, Byeng D. and Kim, Miso},
	urldate = {2025-10-06},
	date = {2021-09-16},
}

@article{SuperCell_1_zhang_topological_2022,
	title = {Topological multipolar corner state in a supercell metasurface and its interplay with two-dimensional materials},
	volume = {10},
	rights = {© 2022 Chinese Laser Press},
	issn = {2327-9125},
	url = {https://opg.optica.org/prj/abstract.cfm?uri=prj-10-4-855},
	doi = {10.1364/PRJ.443025},
	pages = {855--869},
	number = {4},
	journaltitle = {Photonics Research},
	shortjournal = {Photon. Res., {PRJ}},
	author = {Zhang, Zhaojian and Yang, Junbo and Du, Te and Jiang, Xinpeng},
	urldate = {2025-10-06},
	date = {2022-04-01},
	note = {Publisher: Optica Publishing Group},
}

@article{SuperCell_2_chu_temperature_2024,
	title = {Temperature tunability of topological phase transitions and edge states in two-dimensional acoustic topological insulators},
	volume = {14},
	rights = {2024 The Author(s)},
	issn = {2045-2322},
	url = {https://www.nature.com/articles/s41598-024-71021-1},
	doi = {10.1038/s41598-024-71021-1},
	pages = {19793},
	number = {1},
	journaltitle = {Scientific Reports},
	shortjournal = {Sci Rep},
	author = {Chu, Yangyang and Sun, Tong and Wang, Zhaohong and Zhang, Zhifeng},
	urldate = {2025-10-06},
	date = {2024-08-26},
	langid = {english},
	note = {Publisher: Nature Publishing Group},
}

@article{Supercell_3_rajabpoor_alisepahi_breakdown_2023,
	title = {Breakdown of conventional winding number calculation in one-dimensional lattices with interactions beyond nearest neighbors},
	volume = {6},
	rights = {2023 The Author(s)},
	issn = {2399-3650},
	url = {https://www.nature.com/articles/s42005-023-01461-0},
	doi = {10.1038/s42005-023-01461-0},
	pages = {334},
	number = {1},
	journaltitle = {Communications Physics},
	shortjournal = {Commun Phys},
	author = {Rajabpoor Alisepahi, Amir and Sarkar, Siddhartha and Sun, Kai and Ma, Jihong},
	urldate = {2025-07-09},
	date = {2023-11-21},
	langid = {english},
	note = {Publisher: Nature Publishing Group},
}

@article{MetaGen_1_guild_plasmonic-type_2012,
	title = {Plasmonic-type acoustic cloak made of a bilaminate shell},
	volume = {86},
	url = {https://link.aps.org/doi/10.1103/PhysRevB.86.104302},
	doi = {10.1103/PhysRevB.86.104302},
	pages = {104302},
	number = {10},
	journaltitle = {Physical Review B},
	shortjournal = {Phys. Rev. B},
	author = {Guild, Matthew D. and Haberman, Michael R. and Alù, Andrea},
	urldate = {2024-04-16},
	date = {2012-09-07},
	note = {Publisher: American Physical Society},
}

@article{MetaGen_2_zhu_holey-structured_2011,
	title = {A holey-structured metamaterial for acoustic deep-subwavelength imaging},
	volume = {7},
	rights = {2010 Springer Nature Limited},
	issn = {1745-2481},
	url = {https://www.nature.com/articles/nphys1804},
	doi = {10.1038/nphys1804},
	pages = {52--55},
	number = {1},
	journaltitle = {Nature Physics},
	shortjournal = {Nature Phys},
	author = {Zhu, J. and Christensen, J. and Jung, J. and Martin-Moreno, L. and Yin, X. and Fok, L. and Zhang, X. and Garcia-Vidal, F. J.},
	urldate = {2025-07-02},
	date = {2011-01},
	langid = {english},
	note = {Publisher: Nature Publishing Group},
}

@article{MetaGen_3_tuniz_metamaterial_2013,
	title = {Metamaterial fibres for subdiffraction imaging and focusing at terahertz frequencies over optically long distances},
	volume = {4},
	rights = {2013 The Author(s)},
	issn = {2041-1723},
	url = {https://www.nature.com/articles/ncomms3706},
	doi = {10.1038/ncomms3706},
	pages = {2706},
	number = {1},
	journaltitle = {Nature Communications},
	shortjournal = {Nat Commun},
	author = {Tuniz, Alessandro and Kaltenecker, Korbinian J. and Fischer, Bernd M. and Walther, Markus and Fleming, Simon C. and Argyros, Alexander and Kuhlmey, Boris T.},
	urldate = {2025-10-05},
	date = {2013-10-28},
	langid = {english},
	note = {Publisher: Nature Publishing Group},
}

@article{MetaGen_4_zigoneanu_three-dimensional_2014,
	title = {Three-dimensional broadband omnidirectional acoustic ground cloak},
	volume = {13},
	rights = {2014 Springer Nature Limited},
	issn = {1476-4660},
	url = {https://www.nature.com/articles/nmat3901},
	doi = {10.1038/nmat3901},
	pages = {352--355},
	number = {4},
	journaltitle = {Nature Materials},
	shortjournal = {Nature Mater},
	author = {Zigoneanu, Lucian and Popa, Bogdan-Ioan and Cummer, Steven A.},
	urldate = {2025-10-05},
	date = {2014-04},
	langid = {english},
	note = {Publisher: Nature Publishing Group},
}

@article{MetaGen_5_moleron_acoustic_2015,
	title = {Acoustic metamaterial for subwavelength edge detection},
	volume = {6},
	rights = {2015 The Author(s)},
	issn = {2041-1723},
	url = {https://www.nature.com/articles/ncomms9037},
	doi = {10.1038/ncomms9037},
	pages = {8037},
	number = {1},
	journaltitle = {Nature Communications},
	shortjournal = {Nat Commun},
	author = {Molerón, Miguel and Daraio, Chiara},
	urldate = {2025-10-05},
	date = {2015-08-25},
	langid = {english},
	note = {Publisher: Nature Publishing Group},
}

@article{MetaGen_6_zangeneh-nejad_analogue_2021,
	title = {Analogue computing with metamaterials},
	volume = {6},
	rights = {2020 Springer Nature Limited},
	issn = {2058-8437},
	url = {https://www.nature.com/articles/s41578-020-00243-2},
	doi = {10.1038/s41578-020-00243-2},
	pages = {207--225},
	number = {3},
	journaltitle = {Nature Reviews Materials},
	shortjournal = {Nat Rev Mater},
	author = {Zangeneh-Nejad, Farzad and Sounas, Dimitrios L. and Alù, Andrea and Fleury, Romain},
	urldate = {2025-10-05},
	date = {2021-03},
	langid = {english},
	note = {Publisher: Nature Publishing Group},
}

@article{MetaGen_7_zanotto_metamaterial-enabled_2022,
	title = {Metamaterial-enabled asymmetric negative refraction of {GHz} mechanical waves},
	volume = {13},
	rights = {2022 The Author(s)},
	issn = {2041-1723},
	url = {https://www.nature.com/articles/s41467-022-33652-8},
	doi = {10.1038/s41467-022-33652-8},
	pages = {5939},
	number = {1},
	journaltitle = {Nature Communications},
	shortjournal = {Nat Commun},
	author = {Zanotto, Simone and Biasiol, Giorgio and Santos, Paulo V. and Pitanti, Alessandro},
	urldate = {2025-10-05},
	date = {2022-10-08},
	langid = {english},
	note = {Publisher: Nature Publishing Group},
}

@article{MetaInv_1_xiao_geometric_2015,
	title = {Geometric phase and band inversion in periodic acoustic systems},
	volume = {11},
	rights = {2014 Springer Nature Limited},
	issn = {1745-2481},
	url = {https://www.nature.com/articles/nphys3228},
	doi = {10.1038/nphys3228},
	pages = {240--244},
	number = {3},
	journaltitle = {Nature Physics},
	shortjournal = {Nature Phys},
	author = {Xiao, Meng and Ma, Guancong and Yang, Zhiyu and Sheng, Ping and Zhang, Z. Q. and Chan, C. T.},
	urldate = {2025-10-05},
	date = {2015-03},
	langid = {english},
	note = {Publisher: Nature Publishing Group},
}

@article{MetaInv_2_fu_topological_2007,
	title = {Topological insulators with inversion symmetry},
	volume = {76},
	doi = {10.1103/PhysRevB.76.045302},
	number = {4},
	journaltitle = {Physical Review B},
	shortjournal = {Phys. Rev. B},
	author = {Fu, Liang},
	date = {2007},
}

@article{MetaInv_3_zak_berrys_1989,
	title = {Berry’s phase for energy bands in solids},
	volume = {62},
	doi = {10.1103/PhysRevLett.62.2747},
	pages = {2747--2750},
	number = {23},
	journaltitle = {Physical Review Letters},
	shortjournal = {Phys. Rev. Lett.},
	author = {Zak, J.},
	date = {1989},
}

@article{MetaInvAppln_1_de_ponti_localized_2024,
	title = {Localized topological states beyond Fano resonances via counter-propagating wave mode conversion in piezoelectric microelectromechanical devices},
	volume = {15},
	rights = {2024 The Author(s)},
	issn = {2041-1723},
	url = {https://www.nature.com/articles/s41467-024-53925-8},
	doi = {10.1038/s41467-024-53925-8},
	pages = {9617},
	number = {1},
	journaltitle = {Nature Communications},
	shortjournal = {Nat Commun},
	author = {De Ponti, Jacopo M. and Zhao, Xuanyi and Iorio, Luca and Maggioli, Tommaso and Colangelo, Marco and Davaji, Benyamin and Ardito, Raffaele and Craster, Richard V. and Cassella, Cristian},
	urldate = {2025-05-03},
	date = {2024-11-07},
	langid = {english},
	note = {Publisher: Nature Publishing Group},
}

@article{MetaInvAppln_2_xi_soft-clamped_2025,
	title = {A soft-clamped topological waveguide for phonons},
	volume = {642},
	rights = {2025 The Author(s), under exclusive licence to Springer Nature Limited},
	issn = {1476-4687},
	url = {https://www.nature.com/articles/s41586-025-09092-x},
	doi = {10.1038/s41586-025-09092-x},
	pages = {947--953},
	number = {8069},
	journaltitle = {Nature},
	author = {Xi, Xiang and Chernobrovkin, Ilia and Košata, Jan and Kristensen, Mads B. and Langman, Eric and Sørensen, Anders S. and Zilberberg, Oded and Schliesser, Albert},
	urldate = {2025-07-02},
	date = {2025-06},
	langid = {english},
	note = {Publisher: Nature Publishing Group},
}

@article{MetaInvAppln_3_prabith_nonlinear_2024,
	title = {Nonlinear corner states in a topologically nontrivial kagome lattice},
	volume = {110},
	doi = {10.1103/PhysRevB.110.104307},
	number = {10},
	journaltitle = {Physical Review B},
	shortjournal = {Phys. Rev. B},
	author = {Prabith, K.},
	date = {2024},
}

@article{MetaInvAppln_4_lu_observation_2017,
	title = {Observation of topological valley transport of sound in sonic crystals},
	volume = {13},
	rights = {2016 Springer Nature Limited},
	issn = {1745-2481},
	url = {https://www.nature.com/articles/nphys3999},
	doi = {10.1038/nphys3999},
	pages = {369--374},
	number = {4},
	journaltitle = {Nature Physics},
	shortjournal = {Nature Phys},
	author = {Lu, Jiuyang and Qiu, Chunyin and Ye, Liping and Fan, Xiying and Ke, Manzhu and Zhang, Fan and Liu, Zhengyou},
	urldate = {2025-10-05},
	date = {2017-04},
	langid = {english},
	note = {Publisher: Nature Publishing Group},
}

@article{MetaInvAppln_5_mousavi_topologically_2015,
	title = {Topologically protected elastic waves in phononic metamaterials},
	volume = {6},
	rights = {2015 The Author(s)},
	issn = {2041-1723},
	url = {https://www.nature.com/articles/ncomms9682},
	doi = {10.1038/ncomms9682},
	pages = {8682},
	number = {1},
	journaltitle = {Nature Communications},
	shortjournal = {Nat Commun},
	author = {Mousavi, S. Hossein and Khanikaev, Alexander B. and Wang, Zheng},
	urldate = {2025-10-05},
	date = {2015-11-04},
	langid = {english},
	note = {Publisher: Nature Publishing Group},
}

@inproceedings{Nonlin_5_wang_piezoelectric_2015,
	title = {Piezoelectric nonlinearity in {GaN} Lamb mode resonators},
	url = {https://ieeexplore.ieee.org/document/7181091},
	doi = {10.1109/TRANSDUCERS.2015.7181091},
	eventtitle = {2015 Transducers - 2015 18th International Conference on Solid-State Sensors, Actuators and Microsystems ({TRANSDUCERS})},
	pages = {989--992},
	booktitle = {2015 Transducers - 2015 18th International Conference on Solid-State Sensors, Actuators and Microsystems ({TRANSDUCERS})},
	author = {Wang, Siping and Popa, Laura C. and Weinstein, Dana},
	urldate = {2025-10-15},
	date = {2015-06},
	note = {{ISSN}: 2164-1641},
}

@inproceedings{Nonlin_6_lu_study_2015,
	title = {Study of thermal nonlinearity in lithium niobate-based {MEMS} resonators},
	url = {https://ieeexplore.ieee.org/document/7181345},
	doi = {10.1109/TRANSDUCERS.2015.7181345},
	eventtitle = {2015 Transducers - 2015 18th International Conference on Solid-State Sensors, Actuators and Microsystems ({TRANSDUCERS})},
	pages = {1993--1996},
	booktitle = {2015 Transducers - 2015 18th International Conference on Solid-State Sensors, Actuators and Microsystems ({TRANSDUCERS})},
	author = {Lu, Ruochen and Gong, Songbin},
	urldate = {2025-10-15},
	date = {2015-06},
	note = {{ISSN}: 2164-1641},
}

@article{Nonlin_4_segovia-fernandez_thermal_2013,
	title = {Thermal Nonlinearities in Contour Mode {AlN} Resonators},
	volume = {22},
	issn = {1941-0158},
	url = {https://ieeexplore.ieee.org/document/6507328/},
	doi = {10.1109/JMEMS.2013.2252422},
	pages = {976--985},
	number = {4},
	journaltitle = {Journal of Microelectromechanical Systems},
	author = {Segovia-Fernandez, Jeronimo and Piazza, Gianluca},
	urldate = {2025-08-24},
	date = {2013-08},
}

@article{Nonlin_1_tazzoli_experimental_2012,
	title = {Experimental Investigation of Thermally Induced Nonlinearities in Aluminum Nitride Contour-Mode {MEMS} Resonators},
	volume = {33},
	issn = {1558-0563},
	url = {https://ieeexplore.ieee.org/document/6174439},
	doi = {10.1109/LED.2012.2188491},
	pages = {724--726},
	number = {5},
	journaltitle = {{IEEE} Electron Device Letters},
	author = {Tazzoli, Augusto and Rinaldi, Matteo and Piazza, Gianluca},
	urldate = {2025-08-24},
	date = {2012-05},
}

@article{PowerHand_1_wunnicke_thermal_2009,
	title = {Thermal behavior of {BAW} filters at high {RF} power levels},
	volume = {56},
	issn = {1525-8955},
	url = {https://ieeexplore.ieee.org/document/5307500},
	doi = {10.1109/TUFFC.2009.1359},
	pages = {2686--2692},
	number = {12},
	journaltitle = {{IEEE} Transactions on Ultrasonics, Ferroelectrics, and Frequency Control},
	author = {Wunnicke, Olaf and van der Wel, Paul J. and Strijbos, Remco C. and de Bruijn, Frank},
	urldate = {2025-08-25},
	date = {2009-12},
}

@article{Nonlin_3_miller_nonlinear_2014,
	title = {Nonlinear dynamics in aluminum nitride contour-mode resonators},
	volume = {104},
	issn = {0003-6951},
	url = {https://doi.org/10.1063/1.4861461},
	doi = {10.1063/1.4861461},
	pages = {014102},
	number = {1},
	journaltitle = {Applied Physics Letters},
	shortjournal = {Appl. Phys. Lett.},
	author = {Miller, Nicholas and Piazza, Gianluca},
	urldate = {2025-10-04},
	date = {2014-01-08},
}

@article{Nonlin_2_givois_dynamics_2021,
	title = {Dynamics of piezoelectric structures with geometric nonlinearities: A non-intrusive reduced order modelling strategy},
	volume = {253},
	issn = {00457949},
	url = {https://linkinghub.elsevier.com/retrieve/pii/S0045794921000973},
	doi = {10.1016/j.compstruc.2021.106575},
	shorttitle = {Dynamics of piezoelectric structures with geometric nonlinearities},
	pages = {106575},
	journaltitle = {Computers \& Structures},
	shortjournal = {Computers \& Structures},
	author = {Givois, Arthur and Deü, Jean-François and Thomas, Olivier},
	urldate = {2025-08-25},
	date = {2021-09},
	langid = {english},
}

@inproceedings{PowerHand_2_larson_power_2000,
	title = {Power handling and temperature coefficient studies in {FBAR} duplexers for the 1900 {MHz} {PCS} band},
	volume = {1},
	url = {https://ieeexplore.ieee.org/document/922680/},
	doi = {10.1109/ULTSYM.2000.922680},
	eventtitle = {2000 {IEEE} Ultrasonics Symposium. An International Symposium (Cat. No.00CH37121)},
	pages = {869--874 vol.1},
	booktitle = {2000 {IEEE} Ultrasonics Symposium. Proceedings. An International Symposium (Cat. No.00CH37121)},
	author = {Larson, J.D. and Ruby, J.D. and Bradley, R.C. and Wen, J. and Kok, Shong-Lam and Chien, A.},
	urldate = {2025-08-25},
	date = {2000-10},
	note = {{ISSN}: 1051-0117},
}

@inproceedings{PowerHand_3_van_der_wel_thermal_2009,
	title = {Thermal behaviour and reliability of solidly mounted Bulk Acoustic Wave Duplexers under high power {RF} loads},
	url = {https://ieeexplore.ieee.org/document/5173310/},
	doi = {10.1109/IRPS.2009.5173310},
	eventtitle = {2009 {IEEE} International Reliability Physics Symposium},
	pages = {557--561},
	booktitle = {2009 {IEEE} International Reliability Physics Symposium},
	author = {van der Wel, P.J. and Wunnicke, O. and de Bruijn, F. and Strijbos, R.C.},
	urldate = {2025-08-25},
	date = {2009-04},
	note = {{ISSN}: 1938-1891},
}

@article{MatProp_LNB_ogi_acoustic_2002,
	title = {Acoustic spectroscopy of lithium niobate: Elastic and piezoelectric coefficients},
	volume = {92},
	issn = {0021-8979},
	url = {https://doi.org/10.1063/1.1497702},
	doi = {10.1063/1.1497702},
	shorttitle = {Acoustic spectroscopy of lithium niobate},
	pages = {2451--2456},
	number = {5},
	journaltitle = {Journal of Applied Physics},
	shortjournal = {J. Appl. Phys.},
	author = {Ogi, Hirotsugu and Kawasaki, Yasunori and Hirao, Masahiko and Ledbetter, Hassel},
	urldate = {2025-10-04},
	date = {2002-09-01},
}

@article{MatProp_AlScN_wu_characterization_2018,
	title = {Characterization of Elastic Modulus Across the (Al1–{xScx})N System Using {DFT} and Substrate-Effect-Corrected Nanoindentation},
	volume = {65},
	issn = {1525-8955},
	url = {https://ieeexplore.ieee.org/document/8438334},
	doi = {10.1109/TUFFC.2018.2862240},
	pages = {2167--2175},
	number = {11},
	journaltitle = {{IEEE} Transactions on Ultrasonics, Ferroelectrics, and Frequency Control},
	author = {Wu, Dong and Chen, Yachao and Manna, Sukriti and Talley, Kevin and Zakutayev, Andriy and Brennecka, Geoff L. and Ciobanu, Cristian V. and Constantine, Paul and Packard, Corinne E.},
	urldate = {2025-10-04},
	date = {2018-11},
}

@article{MatProp_AlN_Kazan_Elastic,
author = {Kazan, M. and Moussaed, E. and Nader, R. and Masri, P.},
title = {Elastic constants of aluminum nitride},
journal = {physica status solidi c},
volume = {4},
number = {1},
pages = {204-207},
doi = {https://doi.org/10.1002/pssc.200673503},
url = {https://onlinelibrary.wiley.com/doi/abs/10.1002/pssc.200673503},
eprint = {https://onlinelibrary.wiley.com/doi/pdf/10.1002/pssc.200673503},
year = {2007}
}

@article{CVRG_nm_1_legtenberg_comb-drive_1996,
	title = {Comb-drive actuators for large displacements},
	volume = {6},
	issn = {0960-1317},
	url = {https://doi.org/10.1088/0960-1317/6/3/004},
	doi = {10.1088/0960-1317/6/3/004},
	pages = {320},
	number = {3},
	journaltitle = {Journal of Micromechanics and Microengineering},
	shortjournal = {J. Micromech. Microeng.},
	author = {Legtenberg, Rob and Groeneveld, A. W. and Elwenspoek, M.},
	urldate = {2025-10-03},
	date = {1996-09},
	langid = {english},
}

@inproceedings{PMG_5_davaji_towards_2017,
	title = {Towards a surface and bulk excited {SAW} gyroscope},
	url = {https://ieeexplore.ieee.org/document/8092490},
	doi = {10.1109/ULTSYM.2017.8092490},
	eventtitle = {2017 {IEEE} International Ultrasonics Symposium ({IUS})},
	pages = {1--4},
	booktitle = {2017 {IEEE} International Ultrasonics Symposium ({IUS})},
	author = {Davaji, Benyamin and Pinrod, Visarute and Kulkarni, Shrinidhi and Lal, Amit},
	urldate = {2025-04-13},
	date = {2017-09},
	note = {{ISSN}: 1948-5727},
}

@inproceedings{PMG_6_pinrod_coexisting_2018,
	title = {Coexisting Surface and Bulk Gyroscopic Effects},
	url = {https://ieeexplore.ieee.org/document/8579861},
	doi = {10.1109/ULTSYM.2018.8579861},
	eventtitle = {2018 {IEEE} International Ultrasonics Symposium ({IUS})},
	pages = {1--4},
	booktitle = {2018 {IEEE} International Ultrasonics Symposium ({IUS})},
	author = {Pinrod, Visarute and Davaji, Benyamin and Lal, Amit},
	urldate = {2025-04-13},
	date = {2018-10},
	note = {{ISSN}: 1948-5727},
}

@article{PMG_1_liu_4h_2024,
	title = {4H silicon carbide bulk acoustic wave gyroscope with ultra-high Q-factor for on-chip inertial navigation},
	volume = {3},
	rights = {2024 The Author(s)},
	issn = {2731-3395},
	url = {https://www.nature.com/articles/s44172-024-00234-z},
	doi = {10.1038/s44172-024-00234-z},
	pages = {1--9},
	number = {1},
	journaltitle = {Communications Engineering},
	shortjournal = {Commun Eng},
	author = {Liu, Zhenming and Long, Yaoyao and Wehner, Charlotte and Wen, Haoran and Ayazi, Farrokh},
	urldate = {2025-04-13},
	date = {2024-06-25},
	langid = {english},
	note = {Publisher: Nature Publishing Group},
}

@inproceedings{PMG_8_liu_high-q_2022,
	location = {Hilton Head, South Carolina, {USA}},
	title = {A High-Q Solid Disk Baw Gyroscope In Monocrystalline 4H Silicon-Carbide With Sub-Ppm As-Born Frequency Split},
	isbn = {978-1-940470-04-7},
	url = {https://transducer-research-foundation.org/technical_digests/HiltonHead_2022/hh2022_0075.pdf},
	doi = {10.31438/trf.hh2022.17},
	eventtitle = {2022 Solid-State, Actuators, and Microsystems Workshop},
	pages = {75--78},
	booktitle = {2022 Solid-State, Actuators, and Microsystems Workshop Technical Digest},
	publisher = {Transducer Research Foundation},
	author = {Liu, Zhenming and Lotfi, Ardalan and Hardin, Michael and Ayazi, Farrokh},
	urldate = {2025-04-13},
	date = {2022-06-05},
	langid = {english},
}

@article{PMG_11_tian_toroidal_2024,
	title = {A toroidal {SAW} gyroscope with focused {IDTs} for sensitivity enhancement},
	volume = {10},
	rights = {2024 The Author(s)},
	issn = {2055-7434},
	url = {https://www.nature.com/articles/s41378-024-00658-9},
	doi = {10.1038/s41378-024-00658-9},
	pages = {1--10},
	number = {1},
	journaltitle = {Microsystems \& Nanoengineering},
	shortjournal = {Microsyst Nanoeng},
	author = {Tian, Lu and Zhao, Haitao and Shen, Qiang and Chang, Honglong},
	urldate = {2025-04-13},
	date = {2024-03-15},
	langid = {english},
	note = {Publisher: Nature Publishing Group},
}

@article{CVRG_11_li_004_2018,
	title = {0.04 degree-per-hour {MEMS} disk resonator gyroscope with high-quality factor (510 k) and long decaying time constant (74.9 s)},
	volume = {4},
	rights = {2018 The Author(s)},
	issn = {2055-7434},
	url = {https://www.nature.com/articles/s41378-018-0035-0},
	doi = {10.1038/s41378-018-0035-0},
	pages = {32},
	number = {1},
	journaltitle = {Microsystems \& Nanoengineering},
	shortjournal = {Microsyst Nanoeng},
	author = {Li, Qingsong and Xiao, Dingbang and Zhou, Xin and Xu, Yi and Zhuo, Ming and Hou, Zhanqiang and He, Kaixuan and Zhang, Yongmeng and Wu, Xuezhong},
	urldate = {2025-06-26},
	date = {2018-11-19},
	langid = {english},
	note = {Publisher: Nature Publishing Group},
}

@article{PMG_7_hodjat-shamami_eigenmode_2020,
	title = {Eigenmode operation of piezoelectric resonant gyroscopes},
	volume = {6},
	rights = {2020 The Author(s)},
	issn = {2055-7434},
	url = {https://www.nature.com/articles/s41378-020-00204-3},
	doi = {10.1038/s41378-020-00204-3},
	pages = {108},
	number = {1},
	journaltitle = {Microsystems \& Nanoengineering},
	shortjournal = {Microsyst Nanoeng},
	author = {Hodjat-Shamami, Mojtaba and Ayazi, Farrokh},
	urldate = {2025-06-26},
	date = {2020-11-30},
	langid = {english},
	note = {Publisher: Nature Publishing Group},
}

@article{PMG_3_serrano_substrate-decoupled_2016,
	title = {Substrate-decoupled, bulk-acoustic wave gyroscopes: Design and evaluation of next-generation environmentally robust devices},
	volume = {2},
	rights = {2016 The Author(s)},
	issn = {2055-7434},
	url = {https://www.nature.com/articles/micronano201615},
	doi = {10.1038/micronano.2016.15},
	shorttitle = {Substrate-decoupled, bulk-acoustic wave gyroscopes},
	pages = {16015},
	number = {1},
	journaltitle = {Microsystems \& Nanoengineering},
	shortjournal = {Microsyst Nanoeng},
	author = {Serrano, Diego E. and Zaman, Mohammad F. and Rahafrooz, Amir and Hrudey, Peter and Lipka, Ron and Younkin, Duane and Nagpal, Shin and Jafri, Ijaz and Ayazi, Farrokh},
	urldate = {2025-06-26},
	date = {2016-04-25},
	langid = {english},
	note = {Publisher: Nature Publishing Group},
}

@article{PMG_12_qi_bridging_2024,
	title = {Bridging piezoelectric and electrostatic effects: a novel piezo-{MEMS} pitch/roll gyroscope with sub 10°/h bias instability},
	volume = {10},
	rights = {2024 The Author(s)},
	issn = {2055-7434},
	url = {https://www.nature.com/articles/s41378-024-00773-7},
	doi = {10.1038/s41378-024-00773-7},
	shorttitle = {Bridging piezoelectric and electrostatic effects},
	pages = {160},
	number = {1},
	journaltitle = {Microsystems \& Nanoengineering},
	shortjournal = {Microsyst Nanoeng},
	author = {Qi, Zhenxiang and Wang, Bowen and Zhai, Zhaoyang and Wang, Zheng and Xiong, Xingyin and Yang, Wuhao and Bie, Xiaorui and Wang, Yao and Zou, Xudong},
	urldate = {2025-06-26},
	date = {2024-10-30},
	langid = {english},
	note = {Publisher: Nature Publishing Group},
}

@article{Optical_1_wu_silicon_2018,
	title = {Silicon Integrated Interferometric Optical Gyroscope},
	volume = {8},
	rights = {2018 The Author(s)},
	issn = {2045-2322},
	url = {https://www.nature.com/articles/s41598-018-27077-x},
	doi = {10.1038/s41598-018-27077-x},
	pages = {8766},
	number = {1},
	journaltitle = {Scientific Reports},
	shortjournal = {Sci Rep},
	author = {Wu, Beibei and Yu, Yu and Xiong, Jiabi and Zhang, Xinliang},
	urldate = {2025-06-26},
	date = {2018-06-08},
	langid = {english},
	note = {Publisher: Nature Publishing Group},
}

@article{PMG_2_tabrizian_high-frequency_2013,
	title = {High-Frequency {AlN}-on-Silicon Resonant Square Gyroscopes},
	volume = {22},
	issn = {1941-0158},
	url = {https://ieeexplore.ieee.org/document/6566045/},
	doi = {10.1109/JMEMS.2013.2273031},
	pages = {1007--1009},
	number = {5},
	journaltitle = {Journal of Microelectromechanical Systems},
	author = {Tabrizian, Roozbeh and Hodjat-Shamami, Mojtaba and Ayazi, Farrokh},
	urldate = {2025-06-26},
	date = {2013-10},
}

@inproceedings{CVRG_6_weinberg_how_2015,
	title = {How to invent (or not invent) the first silicon {MEMS} gyroscope},
	url = {https://ieeexplore.ieee.org/document/7102372/},
	doi = {10.1109/ISISS.2015.7102372},
	eventtitle = {2015 {IEEE} International Symposium on Inertial Sensors and Systems ({INERTIAL})},
	pages = {1--5},
	booktitle = {2015 {IEEE} International Symposium on Inertial Sensors and Systems ({INERTIAL}) Proceedings},
	author = {Weinberg, Marc S.},
	urldate = {2025-06-26},
	date = {2015-03},
}

@article{Optical_2_wu_mode-assisted_2019,
	title = {Mode-assisted Silicon Integrated Interferometric Optical Gyroscope},
	volume = {9},
	rights = {2019 The Author(s)},
	issn = {2045-2322},
	url = {https://www.nature.com/articles/s41598-019-49380-x},
	doi = {10.1038/s41598-019-49380-x},
	pages = {12946},
	number = {1},
	journaltitle = {Scientific Reports},
	shortjournal = {Sci Rep},
	author = {Wu, Beibei and Yu, Yu and Zhang, Xinliang},
	urldate = {2025-06-27},
	date = {2019-09-10},
	langid = {english},
	note = {Publisher: Nature Publishing Group},
}

@article{CVRG_7_nitzan_self-induced_2015,
	title = {Self-induced parametric amplification arising from nonlinear elastic coupling in a micromechanical resonating disk gyroscope},
	volume = {5},
	rights = {2015 The Author(s)},
	issn = {2045-2322},
	url = {https://www.nature.com/articles/srep09036},
	doi = {10.1038/srep09036},
	pages = {9036},
	number = {1},
	journaltitle = {Scientific Reports},
	shortjournal = {Sci Rep},
	author = {Nitzan, Sarah H. and Zega, Valentina and Li, Mo and Ahn, Chae H. and Corigliano, Alberto and Kenny, Thomas W. and Horsley, David A.},
	urldate = {2025-07-02},
	date = {2015-03-12},
	langid = {english},
	note = {Publisher: Nature Publishing Group},
}

@article{CVRG_5_ahn_encapsulated_2014,
	title = {Encapsulated high frequency (235 {kHz}), high-Q (100 k) disk resonator gyroscope with electrostatic parametric pump},
	volume = {105},
	issn = {0003-6951},
	url = {https://doi.org/10.1063/1.4904468},
	doi = {10.1063/1.4904468},
	pages = {243504},
	number = {24},
	journaltitle = {Applied Physics Letters},
	shortjournal = {Applied Physics Letters},
	author = {Ahn, C. H. and Nitzan, S. and Ng, E. J. and Hong, V. A. and Yang, Y. and Kimbrell, T. and Horsley, D. A. and Kenny, T. W.},
	urldate = {2025-07-02},
	date = {2014-12-16},
}

@article{CVRG_9_zhou_mitigating_2017,
	title = {Mitigating Thermoelastic Dissipation of Flexural Micromechanical Resonators by Decoupling Resonant Frequency from Thermal Relaxation Rate},
	volume = {8},
	url = {https://link.aps.org/doi/10.1103/PhysRevApplied.8.064033},
	doi = {10.1103/PhysRevApplied.8.064033},
	pages = {064033},
	number = {6},
	journaltitle = {Physical Review Applied},
	shortjournal = {Phys. Rev. Appl.},
	author = {Zhou, Xin and Xiao, Dingbang and Wu, Xuezhong and Li, Qingsong and Hou, Zhanqiang and He, Kaixuan and Wu, Yulie},
	urldate = {2025-07-02},
	date = {2017-12-29},
	note = {Publisher: American Physical Society},
}

@article{CVRG_14_ren_automatic_2024,
	title = {An automatic Q-factor matching method for eliminating 77\% of the {ZRO} of a {MEMS} vibratory gyroscope in rate mode},
	volume = {10},
	rights = {2024 The Author(s)},
	issn = {2055-7434},
	url = {https://www.nature.com/articles/s41378-024-00695-4},
	doi = {10.1038/s41378-024-00695-4},
	pages = {67},
	number = {1},
	journaltitle = {Microsystems \& Nanoengineering},
	shortjournal = {Microsyst Nanoeng},
	author = {Ren, Jingbo and Zhou, Tong and Zhou, Yi and Li, Yixuan and Su, Yan},
	urldate = {2025-07-02},
	date = {2024-05-24},
	langid = {english},
	note = {Publisher: Nature Publishing Group},
}

@article{Market_6_yazdi_micromachined_1998,
	title = {Micromachined inertial sensors},
	volume = {86},
	issn = {1558-2256},
	url = {https://ieeexplore.ieee.org/document/704269},
	doi = {10.1109/5.704269},
	pages = {1640--1659},
	number = {8},
	journaltitle = {Proceedings of the {IEEE}},
	author = {Yazdi, N. and Ayazi, F. and Najafi, K.},
	urldate = {2025-07-02},
	date = {1998-08},
}

@article{Market_1_acar_environmentally_2009,
	title = {Environmentally Robust {MEMS} Vibratory Gyroscopes for Automotive Applications},
	volume = {9},
	issn = {1558-1748},
	url = {https://ieeexplore.ieee.org/document/5297817},
	doi = {10.1109/JSEN.2009.2026466},
	pages = {1895--1906},
	number = {12},
	journaltitle = {{IEEE} Sensors Journal},
	author = {Acar, Cenk and Schofield, Adam R. and Trusov, Alexander A. and Costlow, Lynn E. and Shkel, Andrei M.},
	urldate = {2025-07-02},
	date = {2009-12},
}

@article{Market_2_barbour_inertial_2001,
	title = {Inertial sensor technology trends},
	volume = {1},
	issn = {1558-1748},
	url = {https://ieeexplore.ieee.org/document/983473},
	doi = {10.1109/7361.983473},
	pages = {332--339},
	number = {4},
	journaltitle = {{IEEE} Sensors Journal},
	author = {Barbour, N. and Schmidt, G.},
	urldate = {2025-07-02},
	date = {2001-12},
}

@article{Market_5_lane_survey_2010,
	title = {A survey of mobile phone sensing},
	volume = {48},
	issn = {1558-1896},
	url = {https://ieeexplore.ieee.org/document/5560598},
	doi = {10.1109/MCOM.2010.5560598},
	pages = {140--150},
	number = {9},
	journaltitle = {{IEEE} Communications Magazine},
	author = {Lane, Nicholas D. and Miluzzo, Emiliano and Lu, Hong and Peebles, Daniel and Choudhury, Tanzeem and Campbell, Andrew T.},
	urldate = {2025-07-02},
	date = {2010-09},
}

@article{CVRG_4_su_silicon_2014,
	title = {Silicon {MEMS} Disk Resonator Gyroscope With an Integrated {CMOS} Analog Front-End},
	volume = {14},
	issn = {1558-1748},
	url = {https://ieeexplore.ieee.org/document/6858010},
	doi = {10.1109/JSEN.2014.2335735},
	pages = {3426--3432},
	number = {10},
	journaltitle = {{IEEE} Sensors Journal},
	author = {Su, Tsanh-Hung and Nitzan, Sarah H. and Taheri-Tehrani, Parsa and Kline, Mitchell H. and Boser, Bernhard E. and Horsley, David A.},
	urldate = {2025-07-02},
	date = {2014-10},
}

@inproceedings{Market_4_geiger_mems_2008,
	title = {{MEMS} {IMU} for {AHRS} applications},
	url = {https://ieeexplore.ieee.org/document/4569973},
	doi = {10.1109/PLANS.2008.4569973},
	eventtitle = {2008 {IEEE}/{ION} Position, Location and Navigation Symposium},
	pages = {225--231},
	booktitle = {2008 {IEEE}/{ION} Position, Location and Navigation Symposium},
	author = {Geiger, W. and Bartholomeyczik, J. and Breng, U. and Gutmann, W. and Hafen, M. and Handrich, E. and Huber, M. and Jackle, A. and Kempfer, U. and Kopmann, H. and Kunz, J. and Leinfelder, P. and Ohmberger, R. and Probst, U. and Ruf, M. and Spahlinger, G. and Rasch, A. and Straub-Kalthoff, J. and Stroda, M. and Stumpf, K. and Weber, C. and Zimmermann, M. and Zimmermann, S.},
	urldate = {2025-07-02},
	date = {2008-05},
	note = {{ISSN}: 2153-3598},
}

@inproceedings{Market_3_challoner_boeing_2014,
	title = {Boeing Disc Resonator Gyroscope},
	url = {https://ieeexplore.ieee.org/document/6851410},
	doi = {10.1109/PLANS.2014.6851410},
	eventtitle = {2014 {IEEE}/{ION} Position, Location and Navigation Symposium - {PLANS} 2014},
	pages = {504--514},
	booktitle = {2014 {IEEE}/{ION} Position, Location and Navigation Symposium - {PLANS} 2014},
	author = {Challoner, Anthony D. and Ge, Howard H. and Liu, John Y.},
	urldate = {2025-07-02},
	date = {2014-05},
	note = {{ISSN}: 2153-3598},
}

@article{CVRG_2_alper_single-crystal_2005,
	title = {A single-crystal silicon symmetrical and decoupled {MEMS} gyroscope on an insulating substrate},
	volume = {14},
	issn = {1941-0158},
	url = {https://ieeexplore.ieee.org/document/1492422},
	doi = {10.1109/JMEMS.2005.845400},
	pages = {707--717},
	number = {4},
	journaltitle = {Journal of Microelectromechanical Systems},
	author = {Alper, S.E. and Akin, T.},
	urldate = {2025-07-02},
	date = {2005-08},
}

@article{CVRG_1_lee_surfacebulk_2000,
	title = {Surface/bulk micromachined single-crystalline-silicon micro-gyroscope},
	volume = {9},
	issn = {1941-0158},
	url = {https://ieeexplore.ieee.org/document/896779},
	doi = {10.1109/84.896779},
	pages = {557--567},
	number = {4},
	journaltitle = {Journal of Microelectromechanical Systems},
	author = {Lee, Sangwoo and Park, Sangjun and Kim, Jongpal and Lee, Sangchul and Cho, Dong-Il},
	urldate = {2025-07-02},
	date = {2000-12},
}

@article{CVRG_8_wen_resonant_2017,
	title = {Resonant pitch and roll silicon gyroscopes with sub-micron-gap slanted electrodes: Breaking the barrier toward high-performance monolithic inertial measurement units},
	volume = {3},
	rights = {2017 The Author(s)},
	issn = {2055-7434},
	url = {https://www.nature.com/articles/micronano201692},
	doi = {10.1038/micronano.2016.92},
	shorttitle = {Resonant pitch and roll silicon gyroscopes with sub-micron-gap slanted electrodes},
	pages = {16092},
	number = {1},
	journaltitle = {Microsystems \& Nanoengineering},
	shortjournal = {Microsyst Nanoeng},
	author = {Wen, Haoran and Daruwalla, Anosh and Ayazi, Farrokh},
	urldate = {2025-07-02},
	date = {2017-04-24},
	langid = {english},
	note = {Publisher: Nature Publishing Group},
}

@article{PMG_9_erturk_self-aligned_2023,
	title = {Self-aligned single-electrode actuation of tangential and wineglass modes using {PMN}-{PT}},
	volume = {9},
	rights = {2023 The Author(s)},
	issn = {2055-7434},
	url = {https://www.nature.com/articles/s41378-023-00521-3},
	doi = {10.1038/s41378-023-00521-3},
	pages = {52},
	number = {1},
	journaltitle = {Microsystems \& Nanoengineering},
	shortjournal = {Microsyst Nanoeng},
	author = {Erturk, Ozan and Shambaugh, Kilian and Park, Ha-Seong and Lee, Sang-Goo and Bhave, Sunil A.},
	urldate = {2025-07-02},
	date = {2023-05-04},
	langid = {english},
	note = {Publisher: Nature Publishing Group},
}

@article{CVRG_10_zega_new_2018,
	title = {A new {MEMS} three-axial frequency-modulated ({FM}) gyroscope: a mechanical perspective},
	volume = {70},
	issn = {0997-7538},
	url = {https://www.sciencedirect.com/science/article/pii/S0997753817308902},
	doi = {10.1016/j.euromechsol.2018.02.005},
	shorttitle = {A new {MEMS} three-axial frequency-modulated ({FM}) gyroscope},
	pages = {203--212},
	journaltitle = {European Journal of Mechanics - A/Solids},
	shortjournal = {European Journal of Mechanics - A/Solids},
	author = {Zega, Valentina and Comi, Claudia and Minotti, Paolo and Langfelder, Giacomo and Falorni, Luca and Corigliano, Alberto},
	urldate = {2025-07-02},
	date = {2018-07-01},
}

@article{CVRG_13_lysenko_analysis_2021,
	title = {Analysis of frequency response sensor of {MEMS} gyroscope in vacuum chamber},
	volume = {2086},
	issn = {1742-6596},
	url = {https://dx.doi.org/10.1088/1742-6596/2086/1/012197},
	doi = {10.1088/1742-6596/2086/1/012197},
	pages = {012197},
	number = {1},
	journaltitle = {Journal of Physics: Conference Series},
	shortjournal = {J. Phys.: Conf. Ser.},
	author = {Lysenko, I E and Naumenko, D V and Ezhova, O A},
	urldate = {2025-07-02},
	date = {2021-12},
	langid = {english},
	note = {Publisher: {IOP} Publishing},
}

@article{CVRG_12_wu_effects_2021,
	title = {Effects of both the drive- and sense-mode circuit phase delay on {MEMS} gyroscope performance and real-time suppression of the residual fluctuation phase error},
	volume = {31},
	issn = {0960-1317},
	url = {https://dx.doi.org/10.1088/1361-6439/abec1b},
	doi = {10.1088/1361-6439/abec1b},
	pages = {055006},
	number = {5},
	journaltitle = {Journal of Micromechanics and Microengineering},
	shortjournal = {J. Micromech. Microeng.},
	author = {Wu, Haibin and Zheng, Xudong and Wang, Xuetong and Shen, Yaojie and Jin, Zhonghe and Ma, Zhipeng},
	urldate = {2025-07-02},
	date = {2021-04},
	langid = {english},
	note = {Publisher: {IOP} Publishing},
}

@inproceedings{PMG_10_liu_ta_2024,
	title = {{TA} 0.34Deg/Hour Bulk Acoustic Wave Gyroscope in 4H Silicon-Carbide with an Elevated-Temperature Enhanced Q-Factor of 4.6 Million},
	url = {https://ieeexplore.ieee.org/document/10439512},
	doi = {10.1109/MEMS58180.2024.10439512},
	eventtitle = {2024 {IEEE} 37th International Conference on Micro Electro Mechanical Systems ({MEMS})},
	pages = {7--10},
	booktitle = {2024 {IEEE} 37th International Conference on Micro Electro Mechanical Systems ({MEMS})},
	author = {Liu, Zhenming and Long, Yaoyao and Wehner, Charlotte M. and Wen, Haoran and Ayazi, Farrokh},
	urldate = {2025-07-02},
	date = {2024-01},
	note = {{ISSN}: 2160-1968},
}

@inproceedings{CVRG_3_zotov_self-calibrated_2014,
	title = {Self-calibrated {MEMS} gyroscope with {AM}/{FM} operational modes, dynamic range of 180 {dB} and in-run bias stability of 0.1 deg/hr},
	url = {https://ieeexplore.ieee.org/document/7049406},
	doi = {10.1109/InertialSensors.2014.7049406},
	eventtitle = {2014 {DGON} Inertial Sensors and Systems ({ISS})},
	pages = {1--17},
	booktitle = {2014 {DGON} Inertial Sensors and Systems ({ISS})},
	author = {Zotov, S. A. and Prikhodko, I. P. and Simon, B. R. and Trusov, A. A. and Shkel, A. M.},
	urldate = {2025-07-03},
	date = {2014-09},
	note = {{ISSN}: 2377-3480},
}

@article{PMG_4_lee_enhancing_2017,
	title = {Enhancing the sensitivity of three-axis detectable surface acoustic wave gyroscope by using a floating thin piezoelectric membrane},
	volume = {56},
	issn = {1347-4065},
	url = {https://iopscience.iop.org/article/10.7567/JJAP.56.06GN14/meta},
	doi = {10.7567/JJAP.56.06GN14},
	pages = {06GN14},
	number = {6},
	journaltitle = {Japanese Journal of Applied Physics},
	shortjournal = {Jpn. J. Appl. Phys.},
	author = {Lee, Munhwan and Lee, Keekeun},
	urldate = {2025-07-03},
	date = {2017-05-23},
	langid = {english},
	note = {Publisher: {IOP} Publishing},
}

@inproceedings{Coriolis_1_shkel_two_2005,
	title = {Two types of micromachined vibratory gyroscopes},
	url = {https://ieeexplore.ieee.org/document/1597753},
	doi = {10.1109/ICSENS.2005.1597753},
	eventtitle = {2005 {IEEE} {SENSORS}},
	pages = {6 pp.--},
	booktitle = {2005 {IEEE} {SENSORS}},
	author = {Shkel, A.M. and Acar, C. and Painter, C.},
	urldate = {2025-07-14},
	date = {2005-10},
	note = {{ISSN}: 1930-0395},
}

\includepdf[pages=-]{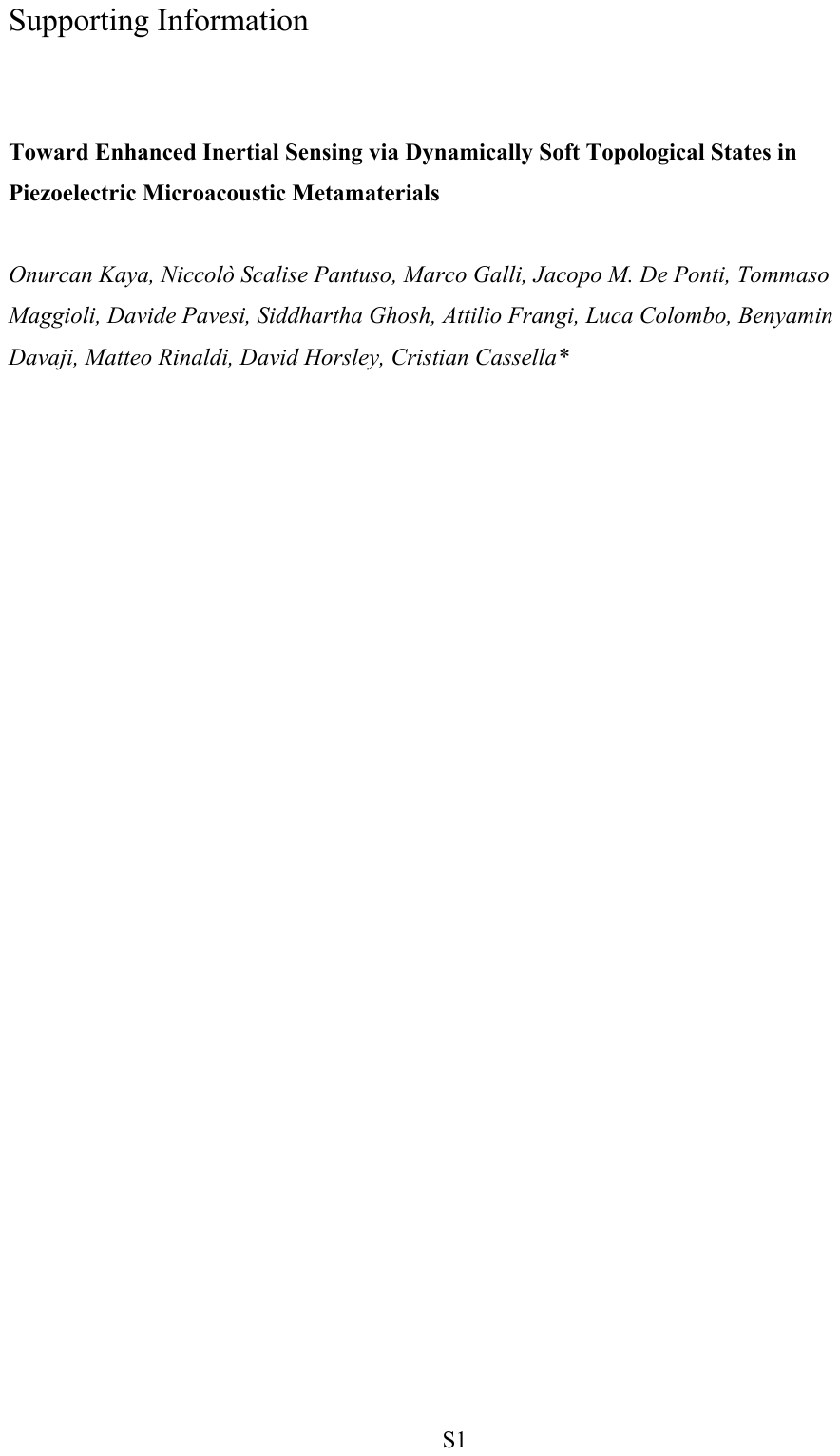}

\end{document}